\documentclass[useAMS,usenatbib,fleqn,a4paper]{mn2e}

\newcommand{\mdeg}{\ensuremath{^{\circ}}}
\usepackage{deluxetable}
\usepackage{subfigure}
\usepackage{graphicx}
\usepackage{amsmath}
\usepackage{amssymb}
\usepackage{bm}
\usepackage{hyperref}
\usepackage{url}
\usepackage{breakurl}
\usepackage{threeparttable}
\usepackage{times}
\title[Deep searches for radio counterparts of gamma-ray pulsars]
{Deep searches for decameter wavelength
pulsed emission from \emph{radio-quiet} gamma-ray pulsars}
\author[Y. Maan and H. A. Aswathappa]
{Yogesh Maan$^{1,2}$\thanks{E-mail: yogesh@rri.res.in; Current Address:
National Centre for Radio Astrophysics, Tata Institute of Fundamental Research,
Pune 411007, India.}
and H. A. Aswathappa$^{1}$\\
$^{1}$Raman Research Institute, Bangalore 560080, India\\
$^{2}$Joint Astronomy Programme, Indian Institute of Science,
Bangalore 560012, India}
\date{}
\begin{document}
\maketitle
\label{firstpage}
\begin{abstract}
We report the results of \emph{(a)} extensive follow-up observations
of the gamma-ray pulsar J1732$-$3131 that has been recently detected
at decameter wavelengths, and \emph{(b)} deep searches for counterparts
of 9 other \emph{radio-quiet} gamma-ray pulsars at 34 MHz, using the
Gauribidanur radio telescope. No periodic signal from J1732$-$3131
could be detected above a detection threshold of $8\sigma$, even
with an effective integration time of more than 40 hours. However,
the average profile obtained by combining data from several epochs, at
a dispersion measure of 15.44~pc~cm$^{-3}$, is found to be consistent
with that from the earlier detection of this pulsar at a confidence
level of $99.2\%$. We present this consistency between the two profiles
as an evidence that
J1732$-$3131 is a faint radio
pulsar with an average flux density of 200--400~mJy at 34~MHz.
Detection sensitivity of our deep searches, despite the extremely
bright sky background at such low frequencies, is generally comparable
to that of higher frequency searches for these pulsars, when scaled
using reasonable assumptions about the underlying pulsar spectrum.
We provide details of our deep searches,
and put stringent upper limits on the decameter wavelength flux
densities of several \emph{radio-quiet} gamma-ray pulsars.
\end{abstract}
\section{Introduction}
The Large Area Telescope (LAT) on board the \textit{Fermi} gamma-ray
satellite, with its unprecedented sensitivity, has revolutionized the
study of gamma-ray emitting pulsars, increasing the known population
from less than 10 to 121 pulsars\footnote{Additional 28 pulsars
detected in gamma-rays are
reported to have publications in preparation \citep{fermi_catalog13},
further increasing the total number of gamma-ray pulsars to 149.}
\citep[][]{fermi_catalog13,Pletsch13}. About one-third (40) of these
pulsars were discovered in blind searches of the LAT data
\citep[][]{Abdo09,Saz10,Pletsch12a,Pletsch12b,Pletsch12c,Pletsch13}.
Despite deep searches at frequencies $\gtrsim500$ MHz
\citep{Saz10,Ray11,Pletsch12a}, confirmed radio counterparts of only
4 of these have been detected so far \citep{Camilo09,Abdo10,Pletsch12a},
suggesting a large fraction of gamma-ray pulsar population to be
\emph{radio-quiet}\footnote{Setting a new convention, the
2$^{\rm nd}$ \textit{Fermi} LAT catalog
of gamma-ray pulsars labels all the pulsars with 1.4 GHz flux density
$<30\,\mu$Jy as ``radio-quiet''. However, as in the usual convention, we
use the term radio-quiet only for those pulsars which have no detectable
radio flux for an observer at earth.}.
\par
A likely explanation for the apparent absence of radio emission from the
majority of the LAT-discovered pulsars is that their narrow radio beams
miss the line of sight towards earth \citep{BJ99,WR11}, and hence appear
as \emph{radio-quiet}. However, the radio emission beam is expected to
become wider at low frequencies \citep[radius-to-frequency mapping in
radio pulsars;][]{Cordes78}, increasing the probability of our line of
sight passing through the beam.
With this in mind, we used the archival data of the pulsar/transient survey
at 34.5 MHz, carried out using the Gauribidanur radio telescope during
2002--2006, to search for decameter-wavelength pulsed emission from
several of the LAT-discovered pulsars. A possible detection of radio
counterpart of the LAT-discovered pulsar J1732$-$3131, resulting from
the above search, was reported earlier \citep*[][hereafter Paper I]{MAD12}.
Weak (and periodic) pulsed emission from J1732$-$3131 was detected in
only one of the several observing sessions.
Although scintillation may explain the detection in only one session,
another likely possibility
is that the radio emission from LAT-discovered pulsars might not
be persistent, i.e., they might appear in \emph{radio-bright} mode
only once in a while. Two categories of radio pulsars --- intermittent
pulsars \citep{Kramer06} and rotating radio transients
\citep[RRATs;][]{McLaughlin06} --- are well known
for such emission behavior.
\subsection{Deep search program: motivation}
Motivated by the intriguing detection of J1732$-$3131, we embarked
on an observing program of deep searches for the decameter-wavelength
counterparts of the so-called radio-quiet gamma-ray pulsars, using the
Gauribidanur radio telescope at 34 MHz. In the first phase of this
\emph{deep search program}, each of the target sources in the selected
sample of 10 gamma-ray pulsars\footnote{Our sample also includes
J1732$-$3131, with the aim of making its confirmatory (re-)detection.}
was observed in multiple ($>20$) sessions. \emph{Deep} searches for
persistent periodic signals were realized by time-aligning and co-adding
the data from these multiple sessions, as described in Section 3.
While the significant enhancement in sensitivity achieved this ways
is important, the deep search program was motivated by two more
crucial factors:
\begin{enumerate}
\item Even for the handful of pulsars which are detectable at such
low frequencies, the received periodic signals are very weak.
Especially at decameter wavelengths, interstellar and
ionospheric scintillation, and contamination from radio frequency
interference (RFI), can hinder detection of such weak signals.
Hence, a weak source, even if intrinsically persistent, may not
be detected in all the observing sessions. In addition, the source may
also be intrinsically variable. Hence, it is important to observe the same
field multiple times.
\item Assuming that our noise statistics are Gaussian, a detection
even at $5\sigma$ might appear quite significant (chance probability
of such a detection is less than $0.6\times10^{-6}$). But the
measured statistics generally deviate from the expected Gaussian nature
due to RFI contamination and/or systematics contributed by the receiver,
and hence the possibility that a $5\sigma$ detection from a single
observing session is due to some weak
RFI can not be ruled out. However, detection of even a relatively weak
periodic signal, but in more than one observing sessions on different days,
consistent in pulse-shape and at the same phase of the period,
is highly unlikely to be a manifestation
of noise (i.e., a chance occurrence) or some RFI. Such consistency across
observing sessions, is therefore crucial to raise the level of confidence
in establishing the astrophysical origin of an otherwise weak signal.
\end{enumerate}
All the LAT-discovered pulsars which we have searched for, are isolated
pulsars with periods in the range 48--444~ms, and only J1813$-$1246 and
J1954$+$2836 have periods below 100~ms. Among the pulsars for which deep
searches have been carried out, J1732$-$3131 is followed-up most extensively
(125 observing sessions). We present here results of our sensitive searches
using these follow-ups of J1732$-$3131 and 9 other pulsars, as well as
those using the archival data,
and provide useful constraints on the decameter-wavelength flux densities
of several radio-quiet gamma-ray pulsars.
Section 2 describes details of the archival data and our new observations.
In section 3, we explain the search methodologies. Section 4 presents
results of follow-up searches of J1732$-$3131 and several other gamma-ray
pulsars, and the upper limits obtained on flux densities of these targets,
followed by conclusions in section 5.
\section[]{Observations and pre-search data processing}
The archival as well as the new observations were carried out using the
Gauribidanur radio telescope. The telescope originally consisted of an array
of 640 dipoles ($160\times4$~rows) in the east-west direction (hereafter
EW array) and an array of 360 dipoles extending southwards from the center
of the EW array \citep*{DSS89}. Presently, \emph{only} the EW arm of this
telescope is maintained, and the survey as well as the new observations
were carried out using this array in coherent phased-array mode. The beam
widths of the EW array are 21 arcmin and $25\mdeg\times\sec{\rm (zenith~angle)}$
in right ascension (RA) and declination (Dec), respectively, with an
effective collective area of about $12000~{\rm m}^2$ at the instrumental
zenith ($+14\mdeg.1$ Dec). The target source is tracked during the
observation by steering the phased-array beam electronically.
In both sets of observations, data were acquired using the portable pulsar
receiver\footnote{\url{http://www.rri.res.in/~dsp_ral/ppr/ppr_main.html}}
(hereafter PPR; Deshpande, Ramkumar, Chandrasekaran and Vinutha, in
preparation) as described in Section 2.3.
\subsection{Survey observations}
The pulsar/transient survey was carried out in the years 2002--2006
using the EW array at 34.5~MHz, with a bandwidth of 1.05~MHz.
The full
accessible declination range ($-45\mdeg$ to $+75\mdeg$) could be
covered with 5 discrete pointings in declination: $-30\mdeg$, $-05\mdeg$,
$+14\mdeg$, $+35\mdeg$ and $+55\mdeg$. Appropriate pointings were
made to cover a large range in right ascension. Apart from J1732$-$3131,
data towards 16 other\footnote{Radio counterparts of 3 of these 16
gamma-ray pulsars are known. The radio counterpart of J1907$+$0602
was reported while our searches were ongoing \citep{Abdo10}, while
those of J1741$-$2054 and J2032$+$4127 were already known \citep{Camilo09}.}
gamma-ray pulsars are available
from single/multiple observing sessions of this survey. Other details
of the survey observations towards these sources are given in
Table~\ref{ulimits_archival}.
\subsection{New observations}
Under the deep search observing program, new observations of 10
radio-quiet gamma-ray pulsars were carried out in multiple sessions
spread over several months in 2012. For these observations,
a bandwidth of 1.53 MHz centered at 34 MHz was used. Further,
these observations could use
only $80\%$ of the potential collecting area, since $20\%$ of the
EW array dipoles ($10\%$ at each of the two far ends) were not available.
However, a slightly larger bandwidth and longer session duration, as
compared to those of the survey observations, together provided about
$18\%$ improvement in sensitivity, despite the $20\%$ loss in the collecting
area. Further relevant details of these observations can be found in
Table~\ref{ulimits_new}. Two radio pulsars, B0834$+$06 and B1919$+$21,
were also observed regularly as ``control pulsars''.
The position coordinates of the pulsars
J0633$+$0632 and J0633$+$1746 ([RA,Dec]$=$[06:34:26,~6.5\mdeg] and
[06:34:38,~17.8\mdeg] respectively, precessed to the epoch of observations)
lie close to each other. We observed both of these pulsars
simultaneously by pointing towards the direction [06:34:26,~10\mdeg.0]
(since both the pulsars fall in the same beam, and well above the half
power points).
\subsection{Data acquisition and pre-search processing}
In each of the observing sessions, PPR
was used to directly record the raw signal voltage sequence at
the Nyquist rate (with 2-bit, 4-level quantization), while tracking
the source. In the off-line processing, the voltage time sequence is
Fourier transformed in blocks of lengths appropriate for a chosen
spectral resolution in the resultant dynamic spectrum, and successive
raw power-spectra are averaged to achieve desired temporal resolution.
For the archival data, appropriate parameters are chosen to achieve
256 spectral channels across 1.05~MHz bandwidth centered around 34.5~MHz,
and a temporal resolution of $\sim1.95$~ms. For the new observations, the
resultant dynamic spectrum consists of 1024 channels across 1.53~MHz bandwidth
centered around 34~MHz, with a temporal resolution of $\sim2$~ms.
\par
To identify RFI contaminated parts of
the data, \emph{robust}\footnote{It is possible that the computed mean
and standard deviation get biased by a few very strong pulses. To get an
unbiased (or robust) estimate, mean and standard deviation are recalculated
by using the previous estimates to detect and exclude the strong pulses
above a given S/N threshold. This process is continued iteratively till
the computed mean and standard deviation no more differ from their
respective values in the previous iteration.} mean and standard deviation
are computed, and an appropriate threshold in signal-to-noise
ratio (S/N) is used separately in the frequency and time domains.
First the RFI contaminated frequency channels
are identified, and data from these channels are excluded while
identifying the time samples contaminated with RFI. The RFI contaminated
frequency channels as well as time samples are excluded from
any further processing.
Most of the observations were conducted in the night time, and typically
only a few percent ($<5\%$) of the data were found to be RFI contaminated.
From the new observations, time-intervals cumulating to about one observing
session duration were rejected for J0633$+$0632/J0633$+$1746 and J1809$-$2332.
Several of the observing sessions towards J1732$-$3131 happened to be in
the day time, and only 85 sessions worth of effective integration time
could be used out of a total of 125 observing sessions.
\section[]{Search methods and sensitivity}
The individual observing session data were searched for presence
of single bright pulses as well as for pulsed signals at the expected
periods of the respective gamma-ray pulsars.
While the detailed methodologies of these two kinds of
searches can be found in \citet{ythesis} and Section~2 of Paper~I,
a brief overview is provided below.
\subsection{Single pulse search}
\par
Searching for bright single pulses involves dedispersing the data
at a number of trial dispersion measures (DM), and subjecting the
individual time series, corresponding to each of the trial DMs,
to a common detection criterion, i.e. an appropriate S/N threshold.
For optimum detections, the individual time series are systematically
smoothed with a template of varying width, effectively carrying out a
search across the pulse-width as well.
We sample the template width range in a logarithmic manner,
with a step of 2 (i.e., we use $2^n$ time-samples wide templates,
where $n$ varies from 0 to a maximum chosen value, in steps of
1)\footnote{As evident from Eq.~\ref{sps_speak}, for a given
peak flux density, the highest achievable S/N of a pulse is directly
proportional to square-root of its width. Hence, for an optimum
width-search, we sample the trial pulse width range in a logarithmic manner.}.
We carried out the single pulse
search in two different ranges of DMs: 0--20~pc~cm$^{-3}$ and
20--50~pc~cm$^{-3}$, with the consecutive trial DMs in the two ranges
differing by 0.01~pc~cm$^{-3}$ and 0.05~pc~cm$^{-3}$, respectively.
The maximum match filter widths used for the two ranges are 128~ms
and 256~ms, respectively. The S/N threshold is chosen based on how
many ``false alarms'' can be tolerated in the final candidate list.
For $N_{tot}$ number of points in a time series, the expected number
of ``false-alarms'', $N_f$, crossing a threshold of $\eta$ (in units of
the rms noise) solely due to noise, are given by:
\begin{equation}
erf(\eta/1.414) = 1 - 2 \times N_f / N_{tot}
\label{eq_thresh}
\end{equation}
where $erf()$ is the \textit{error function}. Allowing 5 false
alarms from each of the trial DMs\footnote{Our choice of tolerable
number of false alarms is admittedly large, to increase the probability
of detecting the faint pulses.}, implies a S/N threshold less than 5.
Note that scaling-up of the denominator on the right-hand side
of Eq.~\ref{eq_thresh} appropriately, so as to account for the number
of trial widths as well, does not make the implied threshold significantly
different from 5.
So, we have used a detection threshold of 5
in our single pulse searches. However, detections
marginally above this threshold can be confirmed only when
reasonable number of single pulses are detected at the same DM.
In case of detection of a single bright pulse, we need to insist
on larger S/N ($\geq8$), so that consistency as well as the dispersive
nature of the signal can be checked across the bandwidth.
\subsection{Search for dispersed periodic pulses}
\par
The periodicity search using data from individual
observing sessions involves folding the time series corresponding
to each of the frequency channels over the expected period of the
respective gamma-ray pulsars. The \emph{folded dynamic spectrum}
is then used to search for a dispersed signal, in a way similar to
that used in the deep searches for dispersed periodic pulses
described below (Section~\ref{sect_dsearch}). We also search over
a narrow range of period offsets around the expected period. Extending
the search in the period-domain is particularly important for the
archival data, since the observation epoch is well before the launch
of the Fermi mission and the validity of the back-projected gamma-ray ephemeris
can not be ensured. For the parameters of our search, the optimum
S/N threshold, as suggested by \citet{Handbook04}, is about 5. However,
we set a slightly higher S/N threshold of 8 to account for any low
level RFI, as well as to be able to check for consistency of a signal
across the observation bandwidth.
\par
The multiple observing sessions towards each of the target sources
allowed us to explore any transient or non-persistent periodic
emission from these pulsars.
The multiple session data from the new observations were
used to carry out deep searches, details of which are given below.
\subsection{Deep search for dispersed periodic pulses}\label{sect_dsearch}
Since the rotation ephemerides for the gamma-ray pulsars are known from
timing of the LAT data\footnote{{\sloppy The up-to-date timing
models of several gamma-ray pulsars are provided by the LAT team at
\url{https://confluence.slac.stanford.edu/display/GLAMCOG/LAT+Gamma-ray+Pulsar+Timing+Models}}},
multiple session data from the new observations could be used
advantageously to enhance our sensitivity for detecting a periodic signal.
For each of our target gamma-ray pulsars, we use the pulsar timing software
\textsc{Tempo}\footnote{For more information about \textsc{Tempo}, please
refer to the website: \url{http://www.atnf.csiro.au/research/pulsar/tempo/} .}
along with the corresponding timing model,
to predict the pulsar period and the pulse phase.
The dynamic spectrum for each of the observing sessions is folded over
the predicted pulse period (the time-ranges identified as RFI-contaminated
in the pre-search processing are excluded). The \emph{folded dynamic spectra}
from all the observing sessions of a particular source are then phase-aligned
and co-added. While co-adding, the average band-shape modulation is
removed, and the frequency channels identified as RFI contaminated in
individual observing sessions are excluded. Also, to account for possible
differences in the effective integration time of individual sessions (i.e.,
the RFI-free observation duration), a suitably weighted average of the
folded dynamic spectra is computed.
\par
To search for a dispersed signal, the final co-added (or more precisely,
\emph{averaged})
folded dynamic spectrum is dedispersed for a number of trial dispersion
measures, and the significance of the resultant average profiles is
assessed. To enhance the signal-to-noise ratio (S/N), the profiles are
smoothed to a resolution of about $20\mdeg$ to $30\mdeg$ in pulse-longitude,
and sum-of-squares (Paper I) or $\chi^2$ \citep{Leahy83} is used as
the figure of merit to assess the profile significance. An in-house
developed software pipeline was used to perform the above search.
The pipeline was successfully verified using observations of our
control pulsars.
%
\begin{figure}
\begin{center}
\includegraphics[width=0.45\textwidth,angle=-90.0]{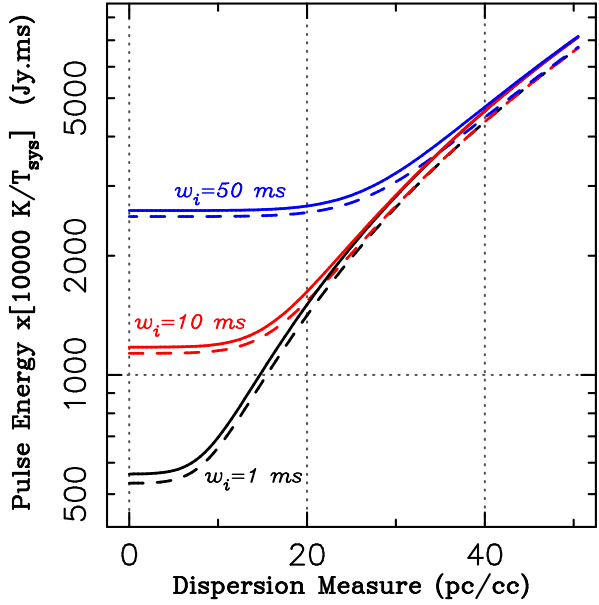}
\caption{The minimum detectable pulse energies, normalized by
$T_{sys}$ in units of $10000\,$K, for intrinsic pulse widths of 1,
10 and 50 ms (the lower, middle and upper pairs of curves, respectively),
are shown as functions of DM. The solid and dashed curves correspond
to the new observations and the archival data, respectively. For pulses
with intrinsic widths smaller than $2$ ms, the observed pulse width is
limited by our sampling time ($\sim2$ ms), and the corresponding
sensitivity curves will nearly follow that for $w_i=1$ ms.
Scatter broadening dependence on DM as modelled by \citet{Bhat04},
and the collecting area corresponding to a pointing declination at
the instrumental zenith has been used.}
\label{fig_sps_smin}
\end{center}
\end{figure}
\begin{table*}
 \centering
  \caption{Searches using the archival data: Observation details and upper flux density limits.}
  \begin{threeparttable}
  \begin{tabular}{@{}rlccccrr@{}}
  \hline
  Sr. & Target PSR\hspace{5mm} & Pointing\hspace{5mm} & Pointing & t$_{obs}$ (s) & $T_{\rm sky}$ & $S_{\rm min}^{\rm SP}$ & $S_{\rm min}^{\rm P0}$ \\
  No. &  & Dec. ($\mdeg$) & offset ($\mdeg$) & [$N_{\rm sessions} \times\tau$] & (K) & (Jy) & (mJy) \\
  \hline
1~ & J0357$+$3205 & $+35$ &  3 & 2$\times$1200 &  18100 &  72 &  218 \\
2~ & J0633$+$0632 & $+14$ &  7 & 1$\times$1200 &  19900 &  91 &  278 \\
3~ & J0633$+$1746 & $+14$ &  4 & 1$\times$1200 &  19900 &  77 &  234 \\
4~ & J1741$-$2054 & $-30$ &  9 & 1$\times$1200 &  62600 & 386 & 1174 \\
5~ & J1809$-$2332 & $-30$ &  6 & 3$\times$1200 &  62400 & 340 & 1036 \\
6~ & J1813$-$1246 & $-05$ &  8 & 4$\times$1200 &  76000 & 387 & 1179 \\
7~ & J1826$-$1256 & $-05$ &  8 & 3$\times$1200 &  80100 & 408 & 1242 \\
8~ & J1846$+$0919 & $+14$ &  5 & 1$\times$1200 &  62900 & 254 &  774 \\
9~ & J1907$+$0602 & $+14$ &  8 & 1$\times$1200 &  71700 & 357 & 1087 \\
10~ & J1954$+$2836 & $+35$ &  6 & 2$\times$1200 &  45000 & 202 &  614 \\
11~ & J1957$+$5033 & $+55$ &  4 & 1$\times$1200 &  35300 & 173 &  527 \\
12~ & J1958$+$2846 & $+35$ &  6 & 2$\times$1200 &  47100 & 211 &  642 \\
13~ & J2021$+$4026 & $+35$ &  5 & 4$\times$1200 &  43200 & 184 &  560 \\
14~ & J2032$+$4127 & $+35$ &  6 & 2$\times$1200 &  38200 & 171 &  521 \\
15~ & J2055$+$2539 &$+35$  &  9 & 2$\times$1200 &  35900 & 199 &  605 \\
16~ & J2238$+$5903 & $+55$ &  4 & 1$\times$1200 &  28100 & 138 &  420 \\
\hline
\label{ulimits_archival}
\end{tabular}
\begin{tablenotes}
\item[] Notes. --- (1) ``Pointing offset'' is the difference between the
pointing declination and the true declination of the target pulsar.
(2) Since the computation of sensitivity limits do not take into
account any possible offset in RA, the limits in some cases might
be underestimated, at most (i.e., in the worst case) by a factor of 2.
(3) $\tau$ is the individual observing session duration, and t$_{obs}$
is the total observation duration of all the sessions towards a particular
source.
\end{tablenotes}
\end{threeparttable}
\end{table*}
\begin{table*}
 \centering
  \caption{Deep searches: Observation details, and comparison of upper flux density limits with those from earlier searches.}
  \begin{threeparttable}
  \begin{tabular}{@{}rlccrrrrcrr@{}}
  \hline
  Sr. & Target PSR & t$_{obs}$ (s) & $T_{\rm sky}$ & $S_{\rm min}^{\rm SP}$ & $S_{\rm min}^{\rm P0}$ & \multicolumn{5}{c}{Comparison with searches at higher frequencies}\\
           \cline{7-11} \noalign{\smallskip}
  No. &  & [$N_{\rm sessions} \times\tau$] & (K) & (Jy) & (mJy) & $S_{\rm previous}$ & $\nu_{\rm obs}$ & Ref. & $S_{\rm previous}^{\rm Scaled}$ & $S_{\rm min}^{\rm P0,Scaled}$ \\
          & & & & & & (mJy) & (MHz) & & ($\mu$Jy) & ($\mu$Jy) \\
  \hline
1~ & J0357$+$3205 &  24$\times$1800 &  17200 &  67 &  34 &   0.043 & 327 & c &   2 &  20 \\
2~ & J0633$+$0632\tnote{$\dagger$} &  45$\times$1800 &  19900 &  77 &  28 &   0.075 & 327 & c &  4 &  17 \\
3~ & J0633$+$1746\tnote{$\dagger$} &  45$\times$1800 &  19900 &  102 &  38 &   150.0 & 35  & a & 94 &  22 \\
4~ & J1732$-$3131\tnote{$\ddagger$} &  85$\times$1800 &  51400 & 271 &  73 &   0.059 & 1374 & c &  57 &  43 \\
5~ & J1809$-$2332 &  20$\times$1800 &  74900 & 348 & 193 &   0.026 & 1352 & c & 24 &  114 \\
6~ & J1836$+$5925 &  33$\times$1800 &  24700 & 128 &  55 &   0.070 & 350 & c &  4 &  32 \\
7~ & J2021$+$4026 &  24$\times$1800 &  44100 & 181 &  92 &   0.051 & 820 & c & 17 &  54 \\
8~ & J2055$+$2539 &  22$\times$1800 &  30400 &  114 &  60 &   0.085 & 327 & b &  5 &  35 \\
9~ & J2139$+$4716 &  23$\times$1800 &  32500 & 143 &  74 &   0.171 & 350 & d & 11 &  44 \\
10~ & J2238$+$5903 &  22$\times$1800 &  29700 & 154 &  82 &   0.027 & 820 & c &  9 &  48 \\
\hline
\label{ulimits_new}
\end{tabular}
\begin{tablenotes}
\item[]{$N_{\rm sessions}$ is modified (lowered) so that t$_{obs}$
provides the \emph{effective} integration time (i.e., the integration
time after excluding the RFI contaminated time intervals).}
\item[$\dagger$]{
The upper flux density limits presented for these pulsars are
modified by the correction factors for the respective offsets from
the pointing declination.}
\item[$\ddagger$]{Using a pulse duty cycle of $50\%$ (instead of $10\%$)
for J1732$-$3131, as indicated by its average profile, would increase
the corresponding $S_{\rm min}^{\rm P0}$ by a factor of 3.}
\item[]{References --- (a) Ramachandran et al. 1998;
(b) Saz Parkinson et al. 2010; (c) Ray et al. 2011; (d) Pletsch et al. 2012b.}
\end{tablenotes}
\end{threeparttable}
\end{table*}
%
\subsection{Single pulse search sensitivity}
In our single-pulse searches, the peak flux density
of a temporally resolved pulse \citep{CM03}, is given by:
\begin{equation}
S_{\rm peak}^{\rm SP} = (S/N)_{\rm peak} \times
\frac{2 k_B T_{sys}}{A_e(z)\sqrt{n_p\, W\,\Delta\nu}}
\label{sps_speak}
\end{equation}
where, $T_{\rm sys}$ is the system temperature, $A_e(z)$ is the
effective collecting area as a function of zenith-angle ($z$), $\Delta\nu$
is the observation bandwidth, $n_p$ is the number of polarizations (1 for
Gauribidanur telescope), and $(S/N)_{\rm peak}$ is peak signal-to-noise
ratio of the pulse corresponding to a smoothing optimum for its observed
width of $W$.
\par
Note that the observed pulse width is contributed to by various pulse
broadening effects, viz. intrinsic pulse width, interstellar scattering,
receiver filter response time, and residual dispersion smearing across
individual frequency channels.
However, the scatter broadening at such low frequencies dominates
over other pulse broadening effects even for moderate values of DM.
Hence, at moderately high DMs, sensitivity of our single pulse search,
in terms of pulse energy (i.e., $S_{\rm peak}^{\rm SP}\times W$),
becomes independent of the \emph{intrinsic} pulse width. This is clearly seen
in Figure~\ref{fig_sps_smin} which shows the minimum detectable pulse
energy for intrinsic pulse widths of 1, 10 and 50 ms, as a function
of DM. For instance, beyond a DM of about 25~pc~cm$^{-3}$, the minimum
detectable pulse energy for all the pulses with intrinsic widths
$\leq 10$~ms is same. We have followed the
scatter broadening dependence on DM as modelled by \citet{Bhat04}.
Also, we have used the collecting area corresponding to a pointing
declination at or near the instrumental zenith of $14\mdeg.1$. For
a declination away from zenith, the sensitivity will decrease by a
factor of $\sec(z)$.
\subsection{Periodic signal search sensitivity}
For periodicity searches, the minimum detectable flux density
$S_{\rm min}^{\rm P0}$, i.e., at the threshold signal-to-noise
ratio $(S/N)_{\rm min}$, is given by \citep*{Vivek82}:
\begin{equation}
S_{\rm min}^{\rm P0} = (S/N)_{\rm min} \times
       \frac{2 k_B T_{\rm sys}}{A_e(z)\sqrt{n_p\, t_{\rm obs}\, \Delta\nu}}
       \sqrt{\frac{W}{P-W}}
\end{equation}
where, $W$ is the pulse width, $P$ is the pulse period and $t_{\rm obs}$
is the total integration time. For archival data, $t_{\rm obs}$ is equal
to the total observation duration of a single session (i.e., about 1200~$s$).
For new observations, $t_{\rm obs}$ equals the cumulative observation
duration of all the sessions.
%
\section[]{Results and Discussion}
\subsection{Searches using the archival data}
Our searches for bright single pulses as well as for periodic signals
using the archival data did not result in any further detection of
decameter-wavelength counterparts of radio-quiet gamma-ray pulsars.
For the archival data, the upper flux density limits for periodic as
well as single pulse emission are presented in Table~\ref{ulimits_archival}.
To enable easy comparison with the flux density limits at higher radio
frequencies available in literature, generally computed for a detection
limit of $5\,\sigma$, the upper limits presented in Table~\ref{ulimits_archival}
are also computed for a (S/N)$_{\rm min}$ of 5.
For the archival observations, our target sources were generally
offset from the pointing center of the beam.
To calculate the factor by which the gain reduces at
the target source declination, relative to the beam-center declination,
we assume a theoretical beam-gain pattern:
$P(\theta)=[\sin{(\pi D \sin{\theta}/\lambda)}/(\pi D \sin{\theta}/\lambda)]^2$,
where $D=20$~m and $\lambda=8.8$~m.
The flux density limits estimated at the beam center
are then scaled-up using the above correction factors computed for
respective source position offsets\footnote{As explained in Paper-I,
system temperature at the beam center is estimated by computing a
weighted average of sky temperature estimates \citep{DU90} at several
points across the large beam using a theoretical beam-gain pattern.}.
Note that we have carried out the above correction only for
the offsets in declination. Possible offsets in RA are less than 1~minute
(i.e., above the half-power points in the beam-gain pattern).
Whenever archival data are available from multiple sessions, the
offsets in RA are different for different sessions, and generally
the RA offset is negligible at least for one of the sessions. Hence,
Table~\ref{ulimits_archival} presents the sensitivity limits for the
best case when there is no offset in RA, and the limits in some cases
might be
underestimated, at most (i.e., in the worst case) by a factor of 2.
The sensitivity limits for the single pulse search ($S_{\rm min}^{\rm SP}$)
are computed for a nominal pulse width of $100$~ms, while those for the
periodicity search ($S_{\rm min}^{\rm P0}$) are computed for a pulse duty
cycle of $10\%$ and observation-duration of a single observing session,
i.e., $1200$~s.
\begin{figure}
\begin{center}
 \includegraphics[width=0.55\textwidth,angle=-90]{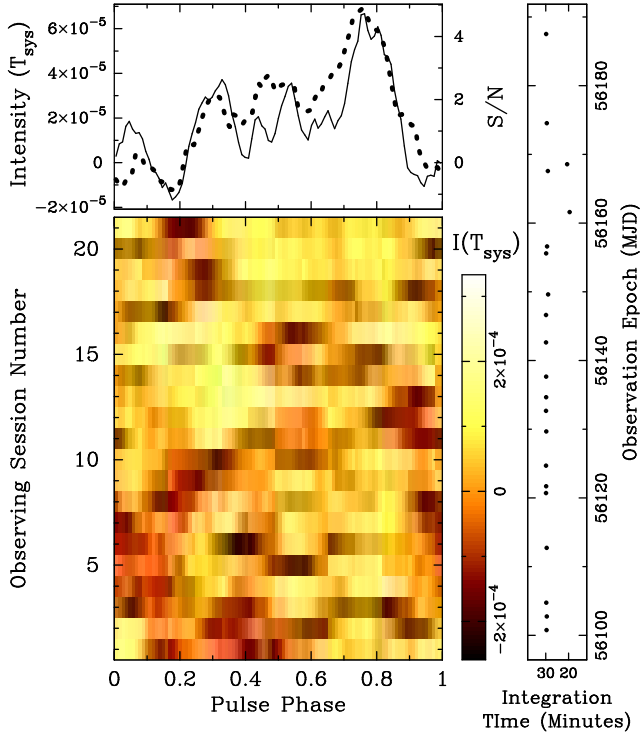}
 \caption[Multi-epoch average radio profiles of J1732$-$3131.]
{Rows in the color image show phase-aligned average profiles of
J1732$-$3131 observed at different epochs (arranged in ascending order
of epoch). The upper panel shows the net average profile (solid line)
and the average profile from our earlier detection using the archival
data (dotted line; see Paper I), for ready comparison. The intensity range
of the latter profile is normalized to that of the former. All the
individual profiles in the color image as well as the two profiles
in the upper panel are smoothed by a 45\mdeg\ wide window. For the
observing sessions corresponding to different rows in the main panel,
the side panel shows the epochs of observation and the effective
integration time (earliest epoch corresponds to first observing session).
Although it is not readily apparent in this figure due to low S/N, the
profile~shape is consistent across the complete range of observing
session~number, i.e., the net average profile has a nearly uniform
contribution from all the sessions (see a moving average filtered
version of the color image in Figure~\ref{supp1}).}
 \label{j1732_mep}
\end{center}
\end{figure}
%
\subsection{Deep follow-up observations of J1732$-$3131}
We carried out extensive follow-up observations of J1732$-$3131,
distributed in 125 sessions, amounting to a total of 62.5 hours of
observation time. In our deep search using an effective integration time
of about 42.5 hours (after rejecting the RFI contaminated time sections),
we could not (re-)detect any readily apparent
(i.e., above a detection threshold of $8\sigma$)
periodic signal from J1732$-$3131.
Our searches for single bright pulses as well as for periodic signal
using the individual session data also did not result in any significant
candidate above our detection threshold of $8\sigma$.
\par
Although we did not have any significant detection, the possibility of a
signal weaker than our detection threshold can not be ruled out. Since
we have an estimate of the DM from our candidate detection of this
pulsar ($15.44\pm0.32$~pc~cm$^{-3}$; Paper I), we can look for
weak periodic signals at this DM that are consistent over multiple
observing sessions.
Furthermore, allowing for the possibility that the periodic signal
might be very weak, if at all present, we carefully chose the
observing sessions that are virtually free from RFI contamination
(assessed by visual inspection of the dynamic spectrum), and where the
dedispersed folded profiles were found to have full-swing S/N (i.e.,
peak-to-peak S/N) more than 4. Such average profiles, corresponding
to 21 sessions, are phase-aligned and presented in Figure~\ref{j1732_mep}.
For comparison, we have overlaid the average profile from the original
detection (dotted line; hereafter \emph{the old profile}) on the net
average profile of all the 21 sessions (solid line; hereafter
\emph{the new profile}) in the upper panel.
The two profiles are manually aligned, since accuracy of the time-stamp
in the archival data is not adequate enough. The two profiles,
observed 10 years apart, exhibit striking similarity, and both are
consistent with each other within the noise uncertainties.
As a quantitative measure of the similarity, the Pearson
(normalized) correlation coefficient between the two profiles is
found to be 0.85.
\par
To further assess statistical significance of the apparent similarity
between the two profiles, we performed Monte-Carlo simulation. An individual
realization in our simulation involves generating a random noise profile and
finding its cross-correlation with the old profile.
To be compatible with the smoothed profiles shown in Figure~\ref{j1732_mep},
the random noise profile is also smoothed with a $45\mdeg$ wide window.
The resultant noise profile is cross-correlated with the old profile at
all possible phase-shifts, and the maximum (normalized) correlation coefficient
is noted down. We simulated 10~million such independent realizations.
The maximum correlation coefficient was found to be $\geq0.85$ (i.e.,
equal to or greater than the correlation found between the old and the
new profile) only in $0.8\%$ of these realizations. Hence, the probability
of the old and the new profiles having the same origin is estimated to
be 0.992. In other words, the two profiles are consistent with each other
at a confidence level of $99.2\%$.
\par
The observed consistency between the average
profile shape obtained by combining data from multiple epochs and that
from the original detection 10 years ago, compels us to infer that
(a)~our candidate detection (Paper I) was not a mere manifestation of noise
or RFI, and hence (b)~the LAT-pulsar J1732$-$3131 is not radio-quiet. If true,
the dispersion measure of this pulsar is $15.44\pm0.32$~pc~cm$^{-3}$
(Paper I). Also, our earlier estimate of the average flux density (i.e.,
pulse-energy/period) of this pulsar in Paper~I ($\sim4$~Jy; at 34.5~MHz)
was most probably affected by scintillation.
The new average profile provides a better estimate (since the scintillation
effects are expected to average out), and suggests the average flux density
to be 200--400~mJy at 34~MHz.
With this new estimate, non-detection of this pulsar
at higher radio frequencies could be explained with a spectral index
$\lesssim -2.3$, assuming no turn-over \citep{Izvekova81} in the spectrum.
This upper limit on the spectral index lies on the steeper
edge of the range of spectral indices for normal pulsars
\citep[$-1.4\pm1.0$;][]{Bates13}.
\subsection{New observations towards other target sources}
In a couple of observing sessions towards the telescope pointing direction
of RA=06:34:26, Dec=10\mdeg, we detected a few ultra-bright pulses at two
different DMs of about $2$~pc~cm$^{-3}$ and $3.3$~pc~cm$^{-3}$, respectively.
However,
when dedispersed at the DMs suggested by the bright single pulses, no
significant signal was found at the expected periodicities of our target
pulsars J0633$+$0632 and J0633$+$1746, which would have been in the telescope
beam centered at above coordinates. Energies of these strong pulses in
the two observing sessions are comparable to typical energies of giant
pulses from the Crab pulsar at decameter wavelengths \citep{Popov06}.
More detailed investigations of these single pulses will be reported
elsewhere.
\par
No significant pulsed (periodic or transient) signal, above a
detection threshold of $8\sigma$, was found towards the directions
of other selected gamma-ray pulsars. The upper limits
on corresponding flux densities, for a detection limit\footnote{As mentioned
earlier, the flux density limits are computed for a (S/N)$_{\rm min}$ of 5,
to enable easy comparison with the flux density limits at higher radio
frequencies available in literature.} of $5\sigma$, are presented in
Table~\ref{ulimits_new}.
For computing the periodic signal search sensitivity ($S_{\rm min}^{\rm P0}$),
we have excluded the time-intervals rejected as RFI contaminated from the
total integration time.
To compare with the earlier searches at higher frequencies, we have also
compiled the flux density limits ($S_{\rm previous}$) from literature,
along with their corresponding observation frequencies ($\nu_{\rm obs}$),
in Table~\ref{ulimits_new}. In case of the limits being available at
several frequencies, the one at the lowest frequency (i.e., closest to 34~MHz)
has been used. Wherever needed, these limits were scaled to $5\sigma$-level,
before compiling into the table. For comparison, our limits at decameter
wavelengths and those from literature are scaled to 1.4 GHz using a spectral
index of $-2.0$, and presented as
$S_{\rm min}^{\rm P0,Scaled}
\left(=S_{\rm min}^{\rm P0}\times\left[\frac{1400}{34}\right]^{-2}\right)$
and $S_{\rm previous}^{\rm Scaled}
\left(=S_{\rm previous}\times\left[\frac{1400}{\nu_{\rm obs}}\right]^{-2}\right)$,
respectively.
We have assumed that there is no spectral turn-over
above our observation frequency (i.e., 34~MHz).
Note that \citet{Bates13} and \citet{Maron00} have estimated the
average spectral index for normal pulsars to be $-1.4\pm1.0$ and
$-1.8\pm0.2$, respectively. Our assumed spectral index (i.e., $-2.0$),
although lying on the steeper side, is well consistent with both these
estimates.
Despite the large background sky-temperature at our observing frequency,
for a couple of pulsars our flux density limits are better than those
from deep searches at higher radio frequencies, and in other cases they
are only within a factor of few of the limits from shorter wavelength
searches
(provided the spectral index of these sources is equal to or
steeper than $-2.0$).
\par
The above comparison of flux density limits may appear to be optimistic,
since we have not assumed any turn-over in the spectrum. However, even
with a turn-over around 80--100~MHz, our flux density limits scale to
typically a few hundreds of $\mu$Jy at 1.4~GHz.
Further, if the lack of radio emission from the LAT-discovered pulsars
is indeed due to unfavorable viewing geometries, then the pulsars which
could possibly be detected at decameter wavelengths can be
expected
to have steep spectra.
If we assume a fairly steep spectrum with an index of $-3.0$ \citep[for
comparison, the spectral index of B0943+10 is $-3.7\pm0.36$;][]{Maron00},
most of our flux limits scale to less than $100\mu$Jy at 1.4~GHz, and
some of them are still comparable to those reported at higher frequencies.
\par
The possibility that some of our target sources are ``radio-loud'',
but have flux densities below our detection limits, can not be ruled
out. The very faint radio emission from J1732$-$3131 which could be
assessed only by making use of its DM estimated from earlier detection
(Paper~I), indicates the possibility of very faint emission from
a few more of the (so far) \emph{radio-quiet} gamma-ray pulsars.
However, lack of radio detection from most of our target sources
indicates that a large fraction of our sample may indeed be radio-quiet.
Consequently, the high fraction of gamma-ray pulsars being radio-quiet
is consistent with the predictions of ``narrow polar-cap'' models
\citep[e.g.,][]{Sturrock71,RS75} for radio beams and ``fan-beam outer
magnetosphere'' models \citep[e.g.,][]{Romani96}
for gamma-ray emission.
%
%
\section[]{Conclusions}
The following points summarize the results of our deep searches for
decameter wavelength counterparts of several \emph{radio-quiet}
gamma-ray pulsars:
\begin{enumerate}
\item We have shown that the 34~MHz average profile of the LAT-discovered
pulsar J1732$-$3131 obtained by effectively integrating over more than
10~hours of new observations carried out at different epochs
(Figure~\ref{j1732_mep}) is consistent with that from the first radio
detection of this pulsar \citep{MAD12} at a confidence level of $99.2\%$.
We present this consistency as an evidence that J1732$-$3131 is a faint
radio pulsar \emph{(and not~radio-quiet)} at decameter wavelengths.
\item We have put stringent upper limits on pulsed (transient as well as
periodic signal) radio emission from several of the radio-quiet gamma-ray
pulsars at decameter wavelengths (Table~\ref{ulimits_new}). Despite the
extremely bright sky background
at decameter wavelengths, the flux density limits obtained from
our deep searches are comparable to those from higher frequency searches
of these pulsars, when scaled to 1.4~GHz assuming a spectral index of
$-2.0$ and no turn-over in the spectrum.
\end{enumerate}
We would also like to emphasize that in the process of
carrying out the deep searches, the Gauribidanur radio telescope is
now appropriately equipped with a sensitive setup to detect and
study known periodic signals with average flux densities
as low as a few mJy, even at such low frequencies.

\section*{Acknowledgments}
We gratefully acknowledge the support from the observatory staff.
We thank the anonymous referee for a critical review of our manuscript,
as well as for the comments and suggestions which helped in improving the
manuscript.
YM is thankful to Avinash Deshpande for useful discussions and
comments on the manuscript. YM is grateful to Paul~Ray and other
members of the LAT-team for providing the up-to-date timing models
of several gamma-ray pulsars. We gratefully thank Indrajit V. Barve,
Hariharan K., Rajalingam M., and several other colleagues, for their
help with the observations at several occasions.
The Gauribidanur radio telescope is jointly operated by
the Raman Research Institute and the Indian Institute of
Astrophysics.

\renewcommand{\thefigure}{S1}
\begin{figure}
\begin{center}
\includegraphics[scale=0.65,angle=-90]{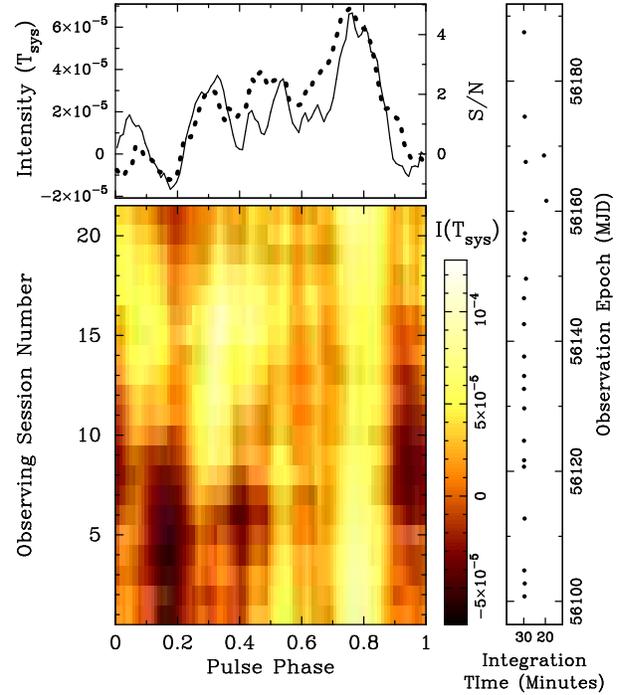}
\caption{Same as Figure $2$ in the main paper, except that each row in the
color image now represents an average profile obtained from 11 observing
sessions --- an average of profiles from $\pm5$ sessions around the
session corresponding to the row. For example, the profile corresponding
to session number 15 represents an average of profiles from the sessions
in the session number range 10--20. Note that, at the edges of the session
number range, the moving average filter is wrapped around. For example, the
$20^{\rm th}$ row profile corresponds to an average of profiles from the
sessions in the ranges 15--21 and 1--4. Note that all the 4 components
(at pulse~phases $\sim0.8$, $\sim0.55$, $\sim0.3$, and $\sim0.08$)
exhibit consistency across the observing sessions, with the brightest
component at pulse~phase $\sim0.8$ \emph{appearing} to be, as expected,
most consistent.}
\label{supp1}
\end{center}
\end{figure}
%
%
%
%
%
\label{lastpage}
\end{document}